\documentclass[%
 superscriptaddress,
showpacs,preprintnumbers,
 amsmath,
 amssymb,
 aps,
 pra,
showkeys,
onecolumn,
notitlepage,
11pt,
]{revtex4-1}
\setcitestyle{authoryear,semicolon,round,,yysep={,}}

\usepackage{graphicx}
\usepackage{dcolumn}
\usepackage{bm}
\usepackage{hyperref}

\usepackage[
text={6in, 8in},centering,
a4paper,
top=1.5in,
bottom=1.5in,
height=10in,a4paper,
]{geometry}

\usepackage{lipsum}

\usepackage{multirow}

\usepackage{xcolor}
\usepackage{soul}
\definecolor{reddish}{HTML}{FBB4AE}
\definecolor{blueish}{HTML}{B3CDE3}
\definecolor{magentish}{HTML}{FF00AA}
\definecolor{greenish}{HTML}{a1d99b}

\usepackage{colortbl}
\definecolor{GrayAbstract}{gray}{0.97}
\definecolor{Gray2}{gray}{0.85}
\definecolor{Gray}{gray}{0.95}

\usepackage{mathptmx}

\usepackage{titlesec}
\titleformat*{\section}{\large\bfseries}
\titleformat*{\subsection}{\bfseries}
\renewcommand{\figurename}{Fig.}
\titlespacing*{\section}{0pt}{1ex plus 1ex minus .5ex}{0.5ex}
\titlespacing*{\subsection}{0pt}{1ex plus 1ex minus .5ex}{0.5ex}
\definecolor{blueish2}{HTML}{0830Bb}
\hypersetup{colorlinks = true,
            linkcolor = blueish2,
            urlcolor  = blueish2,
            citecolor = blueish2,
            anchorcolor = blueish2}
\usepackage{titlesec}

\makeatletter
\renewcommand*{\fnum@figure}{{\normalfont\bfseries \figurename~\thefigure}}
\makeatother

\usepackage{setspace}
\begin{document}

\title{More crime in cities? On the scaling laws of crime and \\the inadequacy of per capita rankings---a cross-country study}

\author{Marcos Oliveira}
\email{moliveira@tuta.io}
\affiliation{Computer Science, University of Exeter, Exeter, United Kingdom\\}
\affiliation{GESIS--Leibniz Institute for the Social Sciences, Cologne, Germany\\}
\begin{abstract}
\centering
\begin{minipage}{4.8in}
\begin{spacing}{1.2}
\noindent
\newline
Crime rates per capita are used virtually everywhere to rank and compare cities. However, their usage relies on a strong linear assumption that crime increases at the same pace as the number of people in a region. In this paper, we demonstrate that using per capita rates to rank cities can produce substantially different rankings from rankings adjusted for population size. We analyze the population--crime relationship in cities across 12 countries and assess the impact of per capita measurements on crime analyses, depending on the type of offense. In most countries, we find that theft increases superlinearly with population size, whereas burglary increases linearly. Our results reveal that per capita rankings can differ from population-adjusted rankings such that they disagree in approximately half of the top 10 most dangerous cities in the data analysed here. Hence, we advise caution when using crime rates per capita to rank cities and recommend evaluating the linear plausibility before doing so. 
\end{spacing}
\end{minipage}

\end{abstract}

\keywords{crime rate, population size, city, urban scaling, ranking, complex systems}
\maketitle

\section*{Introduction}

\label{intro}
In criminology, it is generally accepted that crime occurs more often in more populated regions. In one of the first works of modern criminology, Balbi and Guerry examined the crime distribution across France in 1825, revealing that some areas experienced more crime than others~\citep{balbi1829statistique,Friendly2007}. To compare these areas, they realized the need to adjust for population size and analyzed crime rates  
instead of raw numbers. This method eliminates the \textit{linear} effect of population size on crime numbers and 
has been used to measure crime and compare cities almost everywhere---from academia to news outlets~\citep{nyt2016,independent2016,siegel2011criminology}. However, this approach overlooks the potential \textit{nonlinear} effects of population and, more importantly, exposes our limited understanding of the population--crime relationship.

Though different criminology theories expect a relationship between population size and crime, they tend to disagree on how crime increases with population~\citep{Chamlin2004,Rotolo2006}. These theories predict divergent population effects, such as linear and superlinear crime growth.
Despite these theoretical disputes, however, crime rates per capita are broadly used by assuming that crime increases linearly with the number of people in a region. Crucially, crime rates are often deemed to be a standard means of comparing crime in cities. 


Yet the widespread adoption of crime rates is arguably due more to tradition~\citep{Boivin2013} rather than its ability to remove the effects of population size. Many urban indicators, including crime, have already been shown to increase nonlinearly with population size~\citep{Bettencourt2007}. When we violate the linear assumption and use rates, we deal with quantities that still have population effects, thus introducing an artifactual bias into rankings and analyses. 

Despite this inadequacy, we only have a limited understanding of the impact of nonlinearity on crime rates. Although previous works have investigated population--crime relationships extensively~\citep{Bettencourt2010,Alves2013b,Gomez-Lievano2012,Hanley2016,Chang2019,Yang2019}, they have failed to quantify the impact of nonlinear relationships on rankings and restricted their analyses to either specific offenses or countries. The lack of  comprehensive systematic studies has limited our knowledge on how the linear assumption influences crime analyses and, more critically, has prevented us from better understanding the effect of population on crime.   

In this work, we analyze burglaries and thefts in 12 countries and investigate how crime rates per capita can misrepresent cities in rankings. Instead of assuming that the population--crime relationship is linear, we estimate this relationship from data using probabilistic scaling analysis~\citep{Leitao2016}. We use our estimates to rank cities while adjusting for population size, and we then examine how these rankings differ from rankings based on rates per capita. In our results, we find that the linear assumption is unjustified. We show that using crime rates to rank cities can lead to rankings that considerably differ from rankings adjusted for population size. Finally, our results reveal contrasting growths of burglaries and thefts with population size, implying that different crime dynamics can produce distinct features at the city level. Our work sheds light on the population--crime relationship and suggests caution in using crime rates per capita. 


\section*{Crime and population size}
Different theoretical perspectives predict the emergence of a relationship between population size and crime. Three main criminology theories expect this relationship: structural, social control, and subcultural~\citep{Chamlin2004,Rotolo2006}. In general, these perspectives agree that variations in the number of people in a region have an impact on the way people interact with one another. These theories, however, differ in the types of changes in social interaction and how they can produce a population--crime relationship.

From a \textit{structural} perspective, a higher number of people increases the chances of social interaction, which increases the occurrence of crime. Two distinct rationales can explain such an increase. \cite{Mayhew1976} posit that crime is a product of human contact: more interaction leads to higher chances of individuals being exploited, offended, or harmed. They claim that a larger population size raises the number of  opportunities for interaction at an increasing rate, which would lead to a superlinear crime growth with population size~\citep{Chamlin2004}. In contrast, \cite{blau1977inequality} implies a linear population--crime relationship. He posits that population aggregation reduces spatial distance among individuals, thereby promoting different social associations such as victimization. At the same time, as conflictive association increases, other integrative associations also increase, leading to a linear growth of crime~\citep{Chamlin2004}. Notably, the structural perspective focuses on the \textit{quantitative} consequences of population growth.

The \textit{social control} perspective advocates that changes in population size have a \textit{qualitative} impact on  social relations, which weakens informal social control mechanisms that inhibit crime~\citep{Groff2015}. 
From this perspective, crime relates to two aspect of a population: size and stability. 
A larger population size leads to higher population density and heterogeneity---not only do individuals have more opportunities for social contacts, but they are also often surrounded by strangers~\citep{Wirth1938}.
This situation makes social integration difficult and promotes a high anonymity, which encourages criminal impulses and harms a community's ability to socially constrain misbehavior~\citep{Freudenburg1986,Sampson1986}.
Similarly, from a systemic viewpoint, any change (i.e., increase or decrease) in population size can have an impact on crime numbers~\citep{Rotolo2006}. From this viewpoint, the understanding is that regular and sustained social interactions produce community networks with effective mechanisms of social control~\citep{Bursik1982}. Population instability, however, hinders the construction of such networks. In communities with unstable population size, residents avoid socially investing in their neighborhoods, which hurts community organization and weakens social control, thus increasing misbehavior and crime~\citep{Sampson1988,Miethe1991}. 

Both social-control and structural perspectives solely focus on individuals' interactions without considering their private interests. These perspectives pay little attention to how unconventional interests increase with urbanization~\citep{Fischer1975} and how these interests relate to misbehavior. 

In contrast, the \textit{subcultural} perspective advocates that population concentration brings together individuals with shared interests, which produces private social networks built around these interests, thereby promoting social support for behavioral choices. \cite{Fischer1975} posits that population size has an impact on the creation, diffusion, and intensification of unconventional interests. He proposes that large populations have a sufficient number of people with specific shared interests, thus enabling social interaction and lead to the emergence of subcultures. The social networks surrounding a subculture bring normative expectations that increase the likelihood of misbehavior and crime~\citep{Fischer1995,Fischer1975}. 


These three perspectives---structural, social control, and subcultural---expect that a higher number of people in an area leads to more crime in that area. In the case of cities, we know that population size is indeed a strong predictor of crime~\citep{Bettencourt2007} . The existence of a population--crime relationship implies that we must adjust for population size to analyze crime in cities properly.

\subsection*{Crime rates per capita}
In the literature, the typical solution for removing the effect of population size from crime numbers is to use ratios such as the following:
\begin{equation}
\text{crime rate per capita} = \dfrac{\text{crime}}{\text{\,population\,}}.
\label{eq:crimerate}
\end{equation}
This ratio is often used together with a multiplier that contextualizes the quantity (e.g., crime per $100,000$ inhabitants;~\citealp{Boivin2013}). 
However, even though crime rates are popularly used, they present at least two inadequacies. First, the way in which we define population affects crime rates. The common approach is to use resident population (e.g., census data) to estimate rates, but this practice can distort the picture of crime in a place: crime is not limited to residents~\citep{Gibbs1976}, and cities attract a substantial number of non-residents~\citep{Stults2015}. Instead, researchers suggest using ambient population~\citep{Andresen2006,Andresen2011} and accounting for criminal opportunities, which depends on the type of crime~\citep{CLARKE1984,Harries1981,Boggs1965,Cohen1985}. 

Second, Eq.~(\ref{eq:crimerate}) assumes that the population--crime relationship is linear.
The rationale behind this equation is that we have a relationship of the form
\begin{equation}
\text{crime} \sim \text{population},
\label{eq:crimepop}
\end{equation}
which means that crime can be \textit{linearly} approximated via population.
Given this linear assumption, when we divide crime by population in Eq.~(\ref{eq:crimerate}), we are trying to cancel out the effect of population on crime. This assumption implies that crime increases at the same pace as population growth. However, not all theoretical perspectives agree with this type of growth, and many urban indicators, including crime, have been shown to increase with population size in a nonlinear fashion~\citep{Bettencourt2007}.

\section*{Cities and scaling laws}
Much research has been devoted to understanding urban growth and its impact on indicators such as gross domestic product, total wages, electrical consumption, and crime~\citep{Bettencourt2007,Bettencourt2010,Bettencourt2013,Gomez-Lievano2016}. 
\cite{Bettencourt2007} have found that a city's population size, denoted by $N$, is a strong predictor of its urban indicators, denoted by $Y$, exhibiting the following relationship:
\begin{equation}
\label{eq:scalinglaw}
Y \sim N^\beta.
\end{equation}
This so-called scaling law tells us that, given the size of a city, we expect certain levels of wealth creation, knowledge production, criminality, and other urban aspects. This expectation suggests general processes underlying urban development~\citep{Bettencourt2013hyp} and indicates that regularities exist in cities despite their idiosyncrasies~\citep{oliveira2019spatial}. To understand this scaling and the urban processes better, we can examine the exponent $\beta$, which describes how an urban indicator grows with population size.

\begin{figure}[b!]
	\centering	
	\includegraphics[width=4.5in]{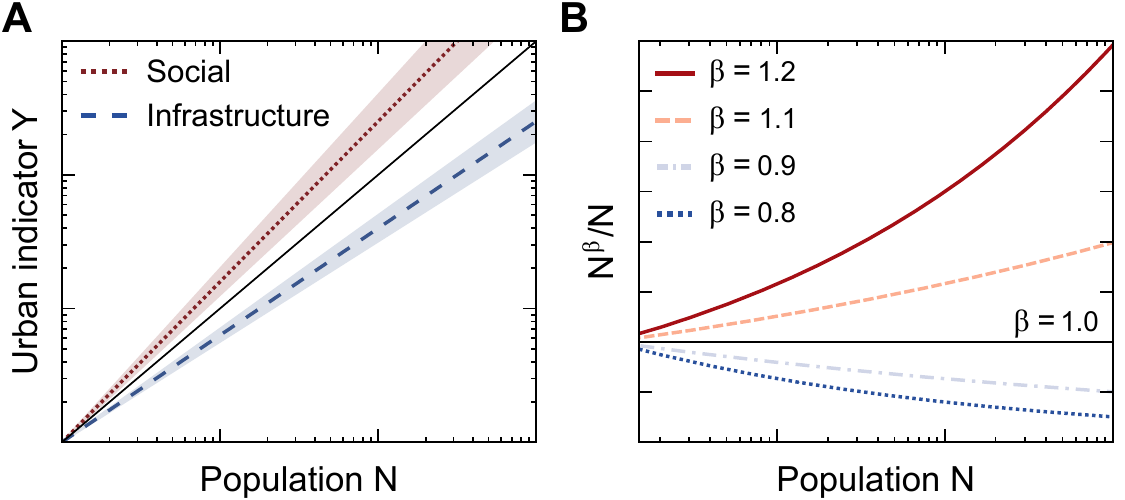}
	\caption{\textbf{Urban scaling laws and rates per capita.} The way in which urban indicators increase with population size depends on the class of the indicator.  (\textbf{A}) Social aspects, such as crime and total wages, increase superlinearly with population size, whereas infrastructural indicators (e.g., road length) increase sublinearly. (\textbf{B})~In nonlinear scenarios, rates per capita still depend on population size.
    }\label{fig1}
\end{figure}

\cite{Bettencourt2007} presented evidence that different categories of urban indicators exhibit distinct growth regimes. They showed that \textit{social} indicators grow faster than \textit{infrastructural} ones~(see Fig.~\ref{fig1}A). Specifically, social indicators, such as the number of patents and total wages, increase superlinearly with population size (i.e., $\beta > 1$), meaning that these indicators grow at an increasing rate with population. In the case of infrastructural aspects (e.g., road surface, length of electrical cables), an economy of scale exists.  As cities grow in population size, these urban indicators increase at a slower pace with $\beta < 1$ (i.e., sublinearly). In both scenarios, because of nonlinearity, we should be careful with per capita analyses.

When we violate the linear assumption of per capita ratios, we deal with quantities that can misrepresent an urban indicator. To demonstrate this, we use Eq.~(\ref{eq:scalinglaw}) to define the per capita rate $C$ of an urban indicator as follows:
\begin{equation*}
\label{eq:scalinglawrate}
    C = \frac{Y}{N} \sim N^{\beta-1},
\end{equation*}
which implies that rates are independent from population only when $\beta$ equals to one---when $\beta \neq 1$, population is not cancelled out from the equation. 
In these nonlinear cases, per capita rates can inflate or deflate the representation of an urban indicator depending on $\beta$ (see Fig.~\ref{fig1}B).
This misrepresentation occurs because population still has an effect on rates.
By definition, we expect that per capita rates are higher in larger cities when $\beta > 1$, whereas when $\beta < 1$, we expect larger cities to have lower rates. When we use rates to compare cities in nonlinear situations, we introduce an artifactual bias. To compare cities properly, previous works have proposed scaled-adjusted indicators that account for population size~\citep{Bettencourt2010,Alves2013b}, supporting the need for population adjustment but failing to quantify the impact of the linear assumption on rankings of urban indicators.

\section*{More crime in cities?}

In the case of crime, researchers have found a superlinear growth with population size. \cite{Bettencourt2007} showed that serious crime in the United States exhibits superlinear scaling with exponent $\beta \approx 1.16$, and some evidence has confirmed similar superlinearity for homicides in Brazil, Colombia, and Mexico~\citep{Gomez-Lievano2012,Alves2013}. Previous works have also found that different kinds of crime in the United Kingdom and in the United States present nonlinear scaling relationships~\citep{Hanley2016,Chang2019,Yang2019}. 
Remarkably, the existence of these scaling laws of crime suggests fundamental urban processes that relate to crime, independent of cities' particularities. 

This regularity manifests itself in the so-called scale-invariance property of scaling laws. It is possible to show that Eq.~(\ref{eq:scalinglaw}) holds the following property: 
\begin{equation}
\label{eq:scaleinvariance}
Y(\kappa N) = g(\kappa)Y(N),
\end{equation}
where $g(\kappa)$ does not depend on $N$~\citep{thurner2018introduction}. From a modeling perspective, this relationship reveals two aspects about crime. First, we can predict crime numbers in cities via a populational  scale transformation $\kappa$~\citep{Bettencourt2013hyp}. This transformation is independent of population size but depends on $\beta$, which tunes the relative increase in crime such that $g(\kappa) = \kappa^\beta$. Second, Eq.~(\ref{eq:scaleinvariance}) implies that crime is present in any city, independent of size. This implication arguably relates to the Durkheimian concept of crime normalcy in that crime is seen as a normal and necessary phenomenon in societies, provided that its numbers are not unusually high~\citep{durkheim1938rules}. Broadly speaking, the scale-invariance property tells us that crime in cities is associated with population in a somewhat predictable fashion. Crucially, this property might give the impression that such regularity is independent of crime type. 


However, different types of crime are connected to social mechanisms in different ways~\citep{Hipp2016} and exhibit unique temporal~\citep{crimeprofiles,Oliveira2018} and spatial characteristics~\citep{Andresen2012,white2014spatial,oliveira2015criminal,Oliveira2017}. It is plausible that the scaling laws of crime depend on crime type. Nevertheless, the literature has mostly focused on either specific countries or crime types. Few studies have systematically examined the scaling of different crime types, and the focus on specific countries has prevented us from better understanding the impact of population on crime. Likewise, the lack of a comprehensive systematic study has limited our knowledge about the impact of the linear assumption on crime rates. We still fail to understand how per capita analyses can misrepresent cities in nonlinear scenarios. 

In this work, we characterize the scaling laws of burglary and theft in 12 countries and investigate how crime rates per capita can misrepresent cities in rankings. Instead of assuming that the population--crime relationship is linear, as described in Eq.~(\ref{eq:crimepop}),  we investigate this relationship under its functional form as follows:
$$\text{crime} \sim f(\text{population}).$$
Specifically, we examine the plausibility of scaling laws to describe the population--crime relationship. To estimate the scaling laws, we use probabilistic scaling analysis, which enables us to characterize the scaling laws of crime. We use our estimates to rank cities while accounting for the effects of population size. Finally, we compare these adjusted rankings with rankings based on per-capita rates (i.e., with the linear assumption).

\section*{Results}

We use data from 12 countries to investigate the relationship between population size and crime at the city level (see the appendix for data sources). Specifically, we examine annual data from Belgium, Canada, Colombia, Denmark, France, Italy, Mexico, Portugal, South Africa, Spain, the United Kingdom, and the United States (see Table~\ref{tab:statscrime_country}). 
In this work, we characterize how crime increases with population size in each country, focusing on burglary and theft. We analyze both crimes in all considered countries, except Mexico, Portugal, and Spain, where we only have data for one type of offense.

\begin{table}[h!]
\centering
	\caption{Burglary and theft annual statistics in 12 countries: number of data points $n$, sample mean $\bar{y}$, sample standard deviation $S$, and maximum value $y_{max}$.\label{tab:statscrime_country}}
   \renewcommand{\arraystretch}{1.1}
   \setlength{\tabcolsep}{2.5pt}
    \begin{tabular}{r|r|rrr|rrr}
    \toprule
    \multicolumn{1}{c|}{\multirow{2}{*}{Country}} & \multicolumn{1}{c|}{\multirow{2}{*}{$n$}} & \multicolumn{3}{c|}{Theft}       & \multicolumn{3}{c}{Burglary}\\ 
    & & \multicolumn{1}{c}{$\bar{y}$} & \multicolumn{1}{c}{$S$} & \multicolumn{1}{c|}{$y_{max}$} & \multicolumn{1}{c}{$\bar{y}$} & \multicolumn{1}{c}{$S$} & \multicolumn{1}{c}{$y_{max}$}\\\hline
Belgium & $588$ 	 & $60.84$ 	& $286.51$ 	& $4,397$ & $95.60$ 	& $209.02$ 	& $2,721$ \\
Canada & $283$ 	 & $1,115.14$ 	& $3,393.88$ 	& $37,150$ & $293.90$ 	& $791.13$ 	& $7,782$ \\
Colombia & $513$ 	 & $182.04$ 	& $1,514.68$ 	& $36,306$ & $40.08$ 	& $228.06$ 	& $4,856$ \\
Denmark & $98$ 	 & $1,157.67$ 	& $3,851.29$ 	& $38,011$ & $330.71$ 	& $330.60$ 	& $2,157$ \\
France & $100$ 	 & $8,311.12$ 	& $12,400.34$ 	& $108,846$ & $2,389.94$ 	& $2,515.24$ 	& $12,511$ \\
Italy & $107$ 	 & $17,470.72$ 	& $30,860.27$ 	& $218,052$ & $2,217.50$ 	& $2,642.61$ 	& $18,101$ \\
Mexico & $1,659$ 	 & $237.56$ 	& $959.59$ 	& $14,999$ & - 	& - 	 & - \\
Portugal & $279$ 	 & - 	& - 	 & - & $51.38$ 	& $86.91$ 	& $850$ \\
South Africa & $199$ 	 & $2,305.23$ 	& $8,758.52$ 	& $93,793$ & $1,190.03$ 	& $3,212.93$ 	& $28,143$ \\
Spain & $144$ 	 & $7,846.72$ 	& $25,111.99$ 	& $236,026$ & - 	& - 	 & - \\
United Kingdom & $313$ 	 & $1,763.43$ 	& $1,965.61$ 	& $19,766$ & $620.98$ 	& $685.40$ 	& $4,825$ \\
United States & $8,337$ 	 & $471.82$ 	& $2,345.27$ 	& $108,376$ & $127.33$ 	& $626.16$ 	& $19,859$ \\\botrule    
    \end{tabular}
\end{table}



\begin{figure}[t!]
	\centering
	{\includegraphics[width=4.7in]{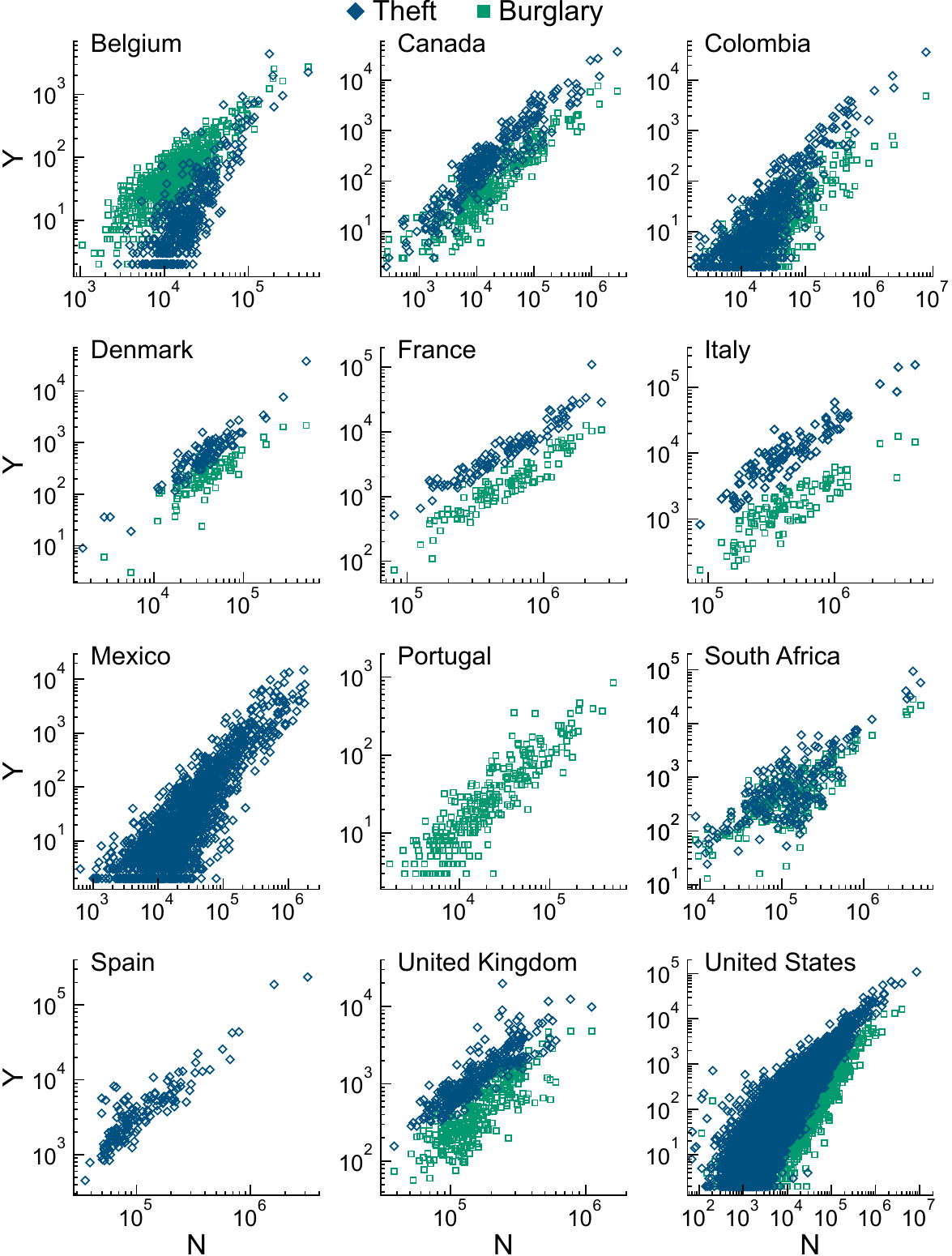}}
	\caption{\textbf{The population--crime relationship in 12 countries.} Different criminology theories expect a relationship between population size and crime, predicting divergent population effects, such as linear and superlinear crime growth. Despite these theoretical disputes, however, crime rates per capita are broadly used by assuming that crime increases linearly with population size. }	\label{fig2}
\end{figure}

\subsection*{The scaling laws of crime in cities}
To assess the relationship between crime $Y$ and population size $N$ (see Fig.~\ref{fig2}), we model ${\rm P}(Y|N)$ using probabilistic scaling analysis (see the Methods section). In our study, we examine whether this relationship follows the general form of $Y \sim N^\beta$. First, we estimate $\beta$ from data, and we then evaluate the plausibility of the model ($p>0.05$)  and the evidence for nonlinearity (i.e., $\beta \neq 1$). Our results reveal that $Y$ and $N$ often exhibit a nonlinear relationship, depending on the type of offense.

\begin{figure}[t!]
	\centering
	\includegraphics[width=4.4in]{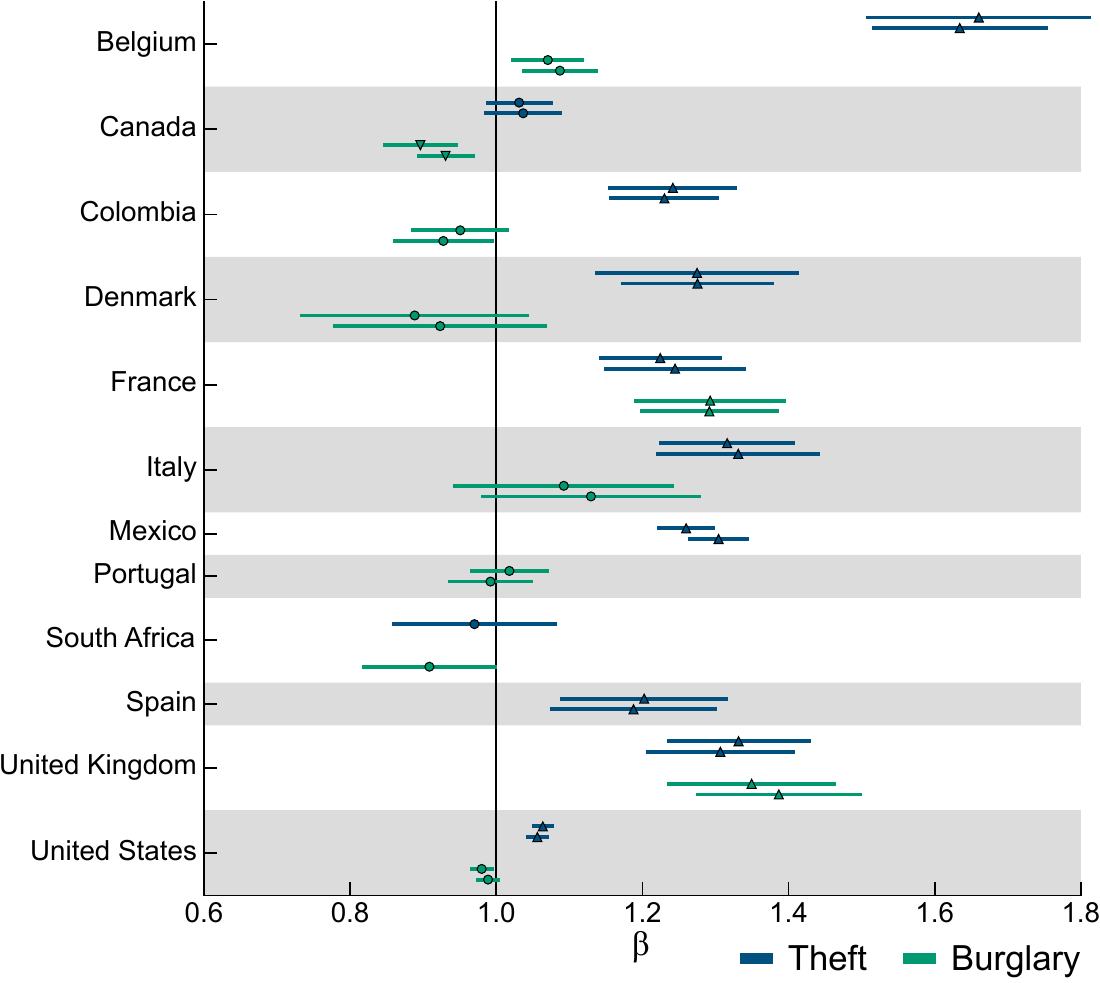}
	\caption{\textbf{The scaling laws of crime.} We find evidence for a nonlinear relationship between crime and population size in more than half of the data sets. In most considered countries, theft exhibits superlinearity, whereas burglary tends to display linearity. In the plot, the lines represent the error bars for the estimated $\beta$ of each country--crime for two consecutive years; circles denote a lack of nonlinearity plausibility; triangles represent superlinearity, and upside-down triangles indicate sublinearity. }
	\label{fig3}
\end{figure}

In most of the considered countries, theft increases with population size superlinearly, whereas burglary tends to increase linearly (see Fig.~\ref{fig3}). Precisely, in 9 out of 11 countries, we find that $\beta$ for theft is above one; our results indicate linearity for theft (i.e., absence of nonlinear plausibility) in Canada and South Africa. In the case of burglary, we are unable to reject linearity in 7 out of 10 countries; in France and the United Kingdom, we find superlinearity, and in Canada, sublinearity. In almost all considered data sets, these estimates are consistent over two consecutive years in the countries for which we have data for different years (see Appendix~I).

Our results suggest  that the general form of $Y\sim N^\beta$ is plausible in most countries, but that this compatibility depends on the offense. 
We find that burglary data are compatible with the \mbox{model ($p>0.05$)} in 80\% of the considered countries. In the case of theft, the superlinear models are compatible with data in five out of nine countries. We note that in Canada and South Africa, where we are unable to reject linearity for theft, the linear model also lacks compatibility with data. 

We find that the estimates of $\beta$ for each offense often have different values across \mbox{countries---for} example, the superlinear estimates of $\beta$ for theft range from $1.10$ to $1.67$. However, when we analyze each country separately, we find that $\beta$ for theft tends to be larger than $\beta$ for burglary in each country, except for France and the United Kingdom.

In summary, we find evidence for a nonlinear relationship between crime and population size in more than half of the considered data sets. Our results indicate that crime often increases with population size at a pace that is different from per capita. This relationship implies that analyses with a linear assumption might create distorted pictures of crime in cities. To understand such distortions, we must examine how nonlinearity influences comparisons of crime in cities, when linearity is assumed.  

\begin{figure}[b!]
	\centering
	{\includegraphics[width=3.57in]{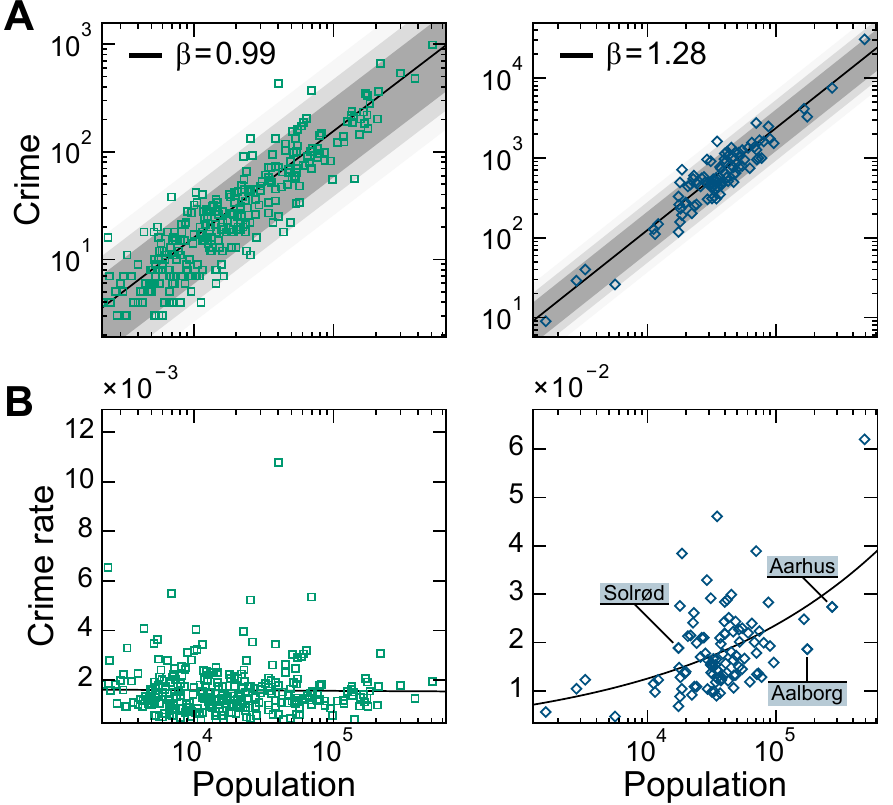}}
	\caption{\textbf{Bias in crime rates per capita.} When crime increases nonlinearly with population size, we have an artifactual bias in crime rates. The linearity in Portugal makes rates independent of size (left). However, in Denmark (right), because of the superlinear growth, we expect larger cities to have higher crime \textit{rates}, but not necessarily more \textit{crime} than expected. For example, though Aalborg and Solr\o{}d have similar theft rates, less crime occurs in Aalborg than expected for cities of the same size, based on the model, whereas Solr\o{}d is above the expectation. }
	\label{fig4}
\end{figure}
\subsection*{The inadequacy of crime rates and per capita rankings}

We investigate how crime rates of the form $C = Y/N$ introduce bias in the comparisons and rankings of cities. To understand this bias, we use Eq.~(\ref{eq:scalinglaw}) to rewrite crime rate as $C \sim N^{\beta - 1}$. This relationship implies that crime rate depends on population size when $\beta \neq 1$. For example, in Portugal and Denmark, this dependency is clear when we analyze burglary and theft numbers (see Fig.~\ref{fig4}). In the case of burglary in Portugal, linearity makes $C$ independent of population size. In Denmark, since theft increases superlinearly, we expect rates to increase with population size. In this country, based on data, the expected theft rate of a small city is lower than the rates of larger cities. 
We must account for this tendency in order to compare crime in cities; otherwise, we introduce bias against larger cities.

To account for the population--crime relationship found in data, we compare cities using the model $P(Y|N)$ as the baseline. We compare the number of crimes in a city with the expectation of the model. For each city $i$ with population size $n_i$, we evaluate the $z$ score of the city with respect to $P(Y|N=n_i)$. The $z$ score indicates how much more or less crime a particular city has in comparison to cities with a similar population size, as expected by the model. These $z$ scores enable us to compare cities in a country and rank them while accounting for population size differences. In contrast, crime rates per capita only adjust for population size in the linear scenario. This approach is similar to previously proposed indicators that adjust for population size~\citep{Bettencourt2010,Alves2013b}. In our case, the adjustment also accounts for the variance. We denote this kind of analysis as a comparison adjusted for the population--crime relationship. 

For example, in Denmark, the theft rate in the municipality of Aalborg ($\approx0.0186$) is almost the same as in Solr\o{}d ($\approx0.0188$). However, less crime occurs in Aalborg than expected for cities of a similar size, while crime in Solr\o{}d is above the model expectation (see Fig.~\ref{fig4}B). This disagreement arises because of the different population sizes. Since Aalborg is more than 10 times larger than Solr\o{}d, we expect rates in Aalborg to be larger than in Solr\o{}d. When we account for this tendency and evaluate their $z$ scores, we find that the $z$ score of Aalborg is $-2.47$, whereas in Solr\o{}d the $z$~score is $2.43$. 


Such inconsistencies have an impact on the crime rankings of cities. The municipality of Aarhus, in Denmark, for example, is ranked among the top 12 cities with the highest theft {rate} in the country. However, when we account for population--crime relationship using $z$ scores, we find that Aarhus is only at the end of the top 54 rankings.

\begin{figure}[b!]
	\centering
	{\includegraphics[width=4.6in]{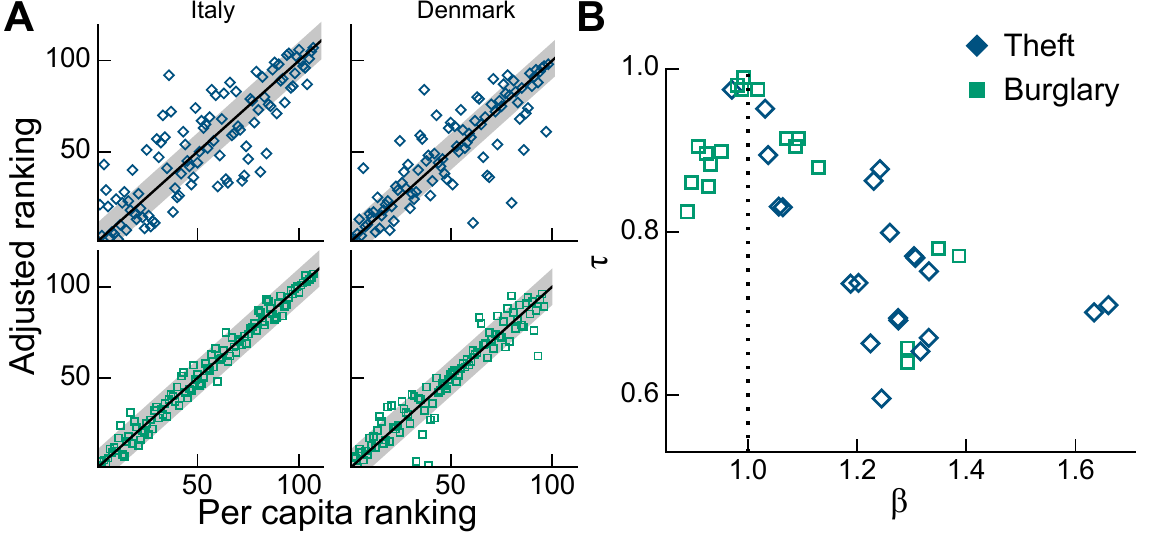}}
	\caption{\textbf{The inadequacy of per capita rankings.} 
Per capita ranking can differ substantially from rankings adjusted for population size, depending on the scaling exponent. In Italy and Denmark, for example, (\textbf{A}) theft ranks (top) diverge considerably more than the ranks for burglary (bottom). Data points represent cities' positions in the rankings. (\textbf{B}) In nonlinear cases, 
these rankings diverge, 
as measured via rank correlation. 
}
	\label{fig5}
\end{figure}

To understand these variations systematically, we compare rankings based on crime rates with rankings that account for the population--crime relationship (i.e., adjusted rankings). Our results reveal that these two rankings create distinct representations of cities. For each considered data set, we rank cities based on their $z$ scores and crime rates $C$, and we then examine the change in the rank of each city. According to our findings, the positions of the cities can change substantially. For instance, in Italy, half of the cities have theft rate ranks that diverge in at least 11 positions from the adjusted ranking~(Fig.~\ref{fig5}A). This disagreement means that these rankings disagree for approximately half of the top 10 most dangerous cities.

We evaluate these discrepancies by using the Kendall rank correlation coefficient $\tau$ to measure the similarity between crime rates and adjusted rankings in the considered countries. We find that these rankings can differ considerably but converge when $\beta \approx 1$. The $\tau$ coefficients for the data sets range from $0.6$ to $1.0$, exhibiting a dependency on the type of crime; or more specifically, on the scaling~(Fig.~\ref{fig5}B). As expected, as $\beta$ approaches $1$, the rankings are more similar to one another. For example, in Italy, in contrast to theft, the burglary rate ranking of half of the cities only differs from the adjusted ranking in a maximum of two positions~(Fig.~\ref{fig5}A). 

\section*{Discussion and Conclusion}
Despite its popularity, comparing cities via crime rates without accounting for population size has a strong assumption that crime increases at the same pace as the number of people in a region. Though previous works have widely investigated the population--crime relationship, they have failed to quantify the impact of nonlinear relationships on rankings and restricted their analyses to either specific offenses or countries. In this work, we analyze crime in different countries to investigate how crime grows with population size and how the widespread assumption of linear growth influences cities' rankings.

First, we analyzed crime in cities from 12 countries to characterize the population--crime relationship statistically, examining the plausibility of scaling laws to describe this relationship. Then, we used our estimates to rank cities and compared how those rankings differ from rankings based on rates per capita.

Our results showed that the assumption of linear crime growth is unfounded. In more than half of the considered data sets, we found evidence for nonlinear crime growth---that is, crime often increases with population size at a different pace than per capita. This nonlinearity introduces a population effect into crime rates, influencing rankings. We demonstrated that using crime rates to rank cities substantially differs from ranking cities adjusted for population size. 

These findings imply that using crime rates per capita---though deemed a standard measure in criminal justice \mbox{statistics---can} create a distorted view of cities' rankings. For example, in superlinear scenarios, we expect larger  cities to have higher crime \textit{rates}. In this case, when we use rates to rank cities, we build rankings whereby large cities are at the top. But, these cities might not experience more \textit{crime} than what we expect from places with a  similar population size. It is an artifactual bias introduced by population effects still present in crime rates. 
 
Such effects arise from nonlinear population effects that persist in rates due to the linear assumption. This assumption is more than just a statistical subtlety. By \textit{assuming} linearity, we essentially overlook cities' context: we ignore the actual impact of population size on crime and how this impact depends on crime type, country, and aggregation units, among other things. For instance, our results indicate that in thefts, linearity is an exception rather than the rule. The indiscriminate use of crime rates neglects significant population--crime interactions that should be considered in order to compare crime in cities properly.

As a result of this inadequacy, we advise caution when using crime rates per capita to compare cities. We recommend evaluating linear plausibility before comparing crime rates. In general, we suggest comparing cities via the $z$ scores computed using the approach~\citep{Leitao2016}  discussed in the manuscript, thereby avoiding crime rates. It is important to emphasize that this inadequacy in rates is relevant only when comparing cities of different population sizes. In analyses without comparisons, a place's crime rate can be seen as a rough indicator that contextualizes crime numbers relative to population size. Additionally, when cities have the same size, comparing crime rates boils down to comparing raw crime numbers. 

In summary, in this work,  we shed light on the population--crime relationship. The linear \textit{assumption} is exhausted and expired. We have resounding evidence of nonlinearity in crime, which disallows us from unjustifiably assuming linearity. In light of our results, we also note that the scaling laws are plausible models only for half of the considered data sets. Better models are thus needed---in particular, models that account for the fact that different crime types relate to population size differently. 
More adequate models will help us better understand the relationship between population and crime.

\section*{Limitations}

Our work presents limitations related to the way in which we define population, crime, and cities. 
First, we note that crime rates depend on how we define population; in our study, we define it as the resident population (i.e., census data). However, crime is not limited to residents~\citep{Gibbs1976}, and cities attract a significant number of non-residents~\citep{Stults2015}. We highlight that this limitation is not specific to our study, and crime rates are often measured using resident population. Previous works have suggested using ambient population and accounting for the number of targets~\citep{Boggs1965,Andresen2006,Andresen2011}. Collecting this data, however, is challenging, especially when dealing with different countries. Future research should investigate crime rates and scaling laws using other definitions of population, particularly using social media data~\citep{Malleson2016,PachecoOM17}.

Second, scaling analyses depend on the definition of what constitutes a city~\citep{Arcaute2014}. In the literature, definitions include legal divisions (e.g., counties, municipalities) and data-driven delineations based on population density and economic interactions~\citep{Cottineau2017}. It is possible that different city definitions yield divergent scaling regimes for the same urban indicator~\citep{Louf2014}. In our work, we only have access to crime data regarding specific aggregation units, and we thus define cities based on official legal divisions by using census data. City definitions in our analysis consequently depend on the country. We emphasize that we investigate whether per capita rankings are justified under a given city definition. Nevertheless, we believe that even though the use of other city definitions might change our quantitative results, our qualitative results are robust: the inadequacy of crime rates is independent of city definitions. When analyzing different definitions of cities, future research should examine scaling divergences as an opportunity to understand the population--crime relationship better.


Finally, cross-national crime analyses have methodological challenges due to  international differences in crime definitions, police and court practices, and reporting rates, among other things~\citep{Takala2008}. Although we avoid direct comparisons of countries' absolute crime numbers in our work, we compare their growth exponents. In this comparison, we assume that cross-national differences have a negligible impact on how crime increases with population, particularly regarding the crime types we analyzed. We understand that some offenses (e.g., sexual assault, drug trafficking) are more sensitive to cross-national  comparisons than the offenses we analyzed here~\citep{Harrendorf2010,Harrendorf2018}. Collecting high-quality international comparative data could help future works in disentangling cross-national differences.


\section*{Methods}



\subsection*{Probabilistic scaling analysis}
We use probabilistic scaling analysis to estimate the scaling laws of crime. Instead of analyzing the linear form of Eq.~(\ref{eq:scalinglaw}), we use the approach developed by \cite{Leitao2016} to estimate the parameters of a distribution $Y|N$ that has the following expectation:
\begin{equation}
\label{eq:scalinglawexpectation}
\mathrm{E}[Y|N] = \lambda N^\beta,
\end{equation}
that is, $N$ scales the expected value of an urban indicator~\citep{Bettencourt2013hyp,Gomez-Lievano2012,Leitao2016}.
Note that this method does not assume that the fluctuations around $\ln y$ and $\ln x$ are normally distributed~\citep{Leitao2016}.
Instead, we compare models for ${\rm P}(Y|N)$ that satisfy the following conditional variance:
\begin{equation}
\label{eq:taylor}
\mathrm{V}[Y|N] = \gamma \mathrm{E}[Y|N]^\delta,
\end{equation}
where typically $\delta \in [1,2]$, since urban systems have been previously shown to exhibit non-trivial fluctuations around the mean---the so-called Taylor's law~\citep{Hanley2014}.
To estimate the scaling laws, we maximize the log-likelihood 
\begin{equation*}
    \mathcal{L} = \ln {\rm P}(y_1, \ldots, y_K|n_1,\ldots,n_K) = 
\sum_{i=1}^K\ln {\rm P}(y_i|n_i),
\end{equation*}
since we assume $y_i$ as an independent realization from ${\rm P}(Y|N)$.
In this work, we use an implementation developed by \cite{Leitao2016} that maximizes the log-likelihood with the ``L-BFGS-B" algorithm. We model  ${\rm P}(Y|N)$ using Gaussian and log-normal distributions in order to analyze whether accounting for the size-dependent variance influences the estimation. 
In the case of  the Gaussian, the conditions from Eq.~(\ref{eq:scalinglawexpectation})  and Eq.~(\ref{eq:taylor}) are satisfied with 
%
%
\begin{equation*}
    \mu_\mathsf{N}(x) = \alpha x^\beta\;\;\;\; \text{and}\;\;\;\; \sigma^2_\mathsf{N}(x) = \gamma(\alpha x^\beta )^\delta,
\end{equation*}
whereas in the case of the log-normal distribution, 
%
\begin{equation*}
    \mu_\mathsf{LN}(x) = \ln \alpha + \beta\ln x - \frac{1}{2} \sigma^2_\mathsf{LN}(x) \;\;\;\; \text{and}\;\;\;\;\sigma^2_\mathsf{LN}(x) = \ln\left[1 + \gamma(\alpha x^\beta )^{\delta - 2}\right].
\end{equation*}
%
%
In the log-normal case, note that, if $\delta = 2$, then the fluctuations are independent of $N$; thus this would be the same as using the minimum least-squares approach~\citep{Leitao2016}.
With this framework, we compare models that have fixed $\delta$ against models wherein  $\delta$ is also included in the optimization process. In the case of the Gaussian, we have fixed $\delta=1$ and free $\delta \in [1,2]$, whereas in the case of the log-normal, we have fixed $\delta=2$ and free $\delta \in [1,3]$. 
In this framework, $p$-values represent a statistic testing two crucial aspects of the modelling: sample independence and model compatibility with data. The statistic consists of the D'Agostino $K^2$ test together with Spearman's rank correlation of residuals, which evaluates compatibility and independence, respectively~\citep{Leitao2016}

Finally, we compare each of the four models individually against the linear alternative (with fixed $\beta = 1$), to test the nonlinearity plausibility.
With the fits of all types of crime and countries, we measure the Bayesian information criterion (${\rm BIC}$), defined as 
\begin{equation*}
    {\rm BIC} = -2\ln \mathcal{L} + k \ln n,
\end{equation*}
where $k$ is the number of free parameters in the model and lower ${\rm BIC}$ values indicate better data description. The ${\rm BIC}$ value of each fit enables us to compare the models' ability to explain data.

\section*{Declarations}

\noindent \textbf{Competing interests}. The authors declare that they have no competing interests.

\noindent \textbf{Funding}. No outside funding was used to support this work.

\noindent \textbf{Author contributions}. All authors read and approved the final manuscript.
    
\noindent \textbf{Availability of data and materials}. All data and source code are available at \url{https://github.com/macoj/scaling_laws_of_crime/}. 



\bibliographystyle{chicago}
\bibliography{main}

\newpage

\section*{Appendices}
\subsection*{Appendix I: Results from the probabilistic scaling analysis}

To test the plausibility of a nonlinear scaling, we compare each model against the linear alternative (i.e., $\beta=1$) using the difference $\Delta {\rm BIC}$ between the fits for each data set. We follow~\cite{Leitao2016} and define three outcomes from this comparison. First, if $\Delta {\rm BIC} < 0$, we say that the model is linear ($\rightarrow$), since we can consider that the linear model explains the data better. Second, if $0 < \Delta {\rm BIC} < 6$, we consider the analysis of $\beta \neq 1$ inconclusive because we do not have enough evidence for the nonlinearity. Finally, if $\Delta {\rm BIC} > 6$, we have evidence in favor of the nonlinear scaling, which can be superlinear ($\nearrow$) or sublinear ($\searrow$). We also use $\Delta {\rm BIC}$ to determine the model ${\rm P}(Y|N)$ that describes the data better. In Table~\ref{tab:results_theft} and Table~\ref{tab:results_burglary}, we summarize the results in that we a dark gray cell indicates the best model based on $\Delta {\rm BIC}$, a light gray cell indicates the best model given a ${\rm P}(Y|N)$ model, and $*$ indicates that the model is plausible ($p$-value$>0.05$).
\begin{table}[b!]
\centering
	\caption{$\beta$ estimates for the case of thefts using log-normal and normal fluctuations.\label{tab:results_theft}}
	\def\arraystretch{1.1}
    \setlength{\tabcolsep}{1.5pt}
    \begin{tabular}{r|ll|ll}
    \toprule
    \multicolumn{1}{c|}{}&\multicolumn{2}{c|}{Log-normal} & \multicolumn{2}{c}{Gaussian} \\
    & \multicolumn{1}{c}{$\delta =2$} & \multicolumn{1}{c|}{$\delta \in [1, 3]$} & \multicolumn{1}{c}{$\delta = 1$} & \multicolumn{1}{c}{$\delta \in [1, 2]$} \\\hline
Belgium (2015) & \cellcolor{Gray}\cellcolor{Gray2}$1.63$ ($0.12$) \parbox[c]{5mm}{\centering$\nearrow$}&$1.64$ ($0.12$) \parbox[c]{5mm}{\centering$\nearrow$}&$2.11$ ($0.27$) \parbox[c]{5mm}{\centering$\nearrow$}&\cellcolor{Gray}$1.67$ ($0.17$) \parbox[c]{5mm}{\centering$\nearrow$}\\
Belgium (2016) & \cellcolor{Gray}\cellcolor{Gray2}$1.66$ ($0.15$) \parbox[c]{5mm}{\centering$\nearrow$}&$1.66$ ($0.14$) \parbox[c]{5mm}{\centering$\nearrow$}&$2.10$ ($0.18$) \parbox[c]{5mm}{\centering$\nearrow$}&\cellcolor{Gray}$1.75$ ($0.19$) \parbox[c]{5mm}{\centering$\nearrow$}\\
Canada (2015) & $1.09$ ($0.06$) \parbox[c]{5mm}{\centering$\nearrow$}&\cellcolor{Gray}\cellcolor{Gray2}$1.04$ ($0.05$) \parbox[c]{5mm}{\centering$\rightarrow$}&$1.07$ ($0.11$) \parbox[c]{5mm}{\centering$\rightarrow$}&\cellcolor{Gray}$1.04$ ($0.06$) \parbox[c]{5mm}{\centering$\rightarrow$}\\
Canada (2016) & \cellcolor{Gray}\cellcolor{Gray2}$1.03$ ($0.04$) \parbox[c]{5mm}{\centering$\rightarrow$}&$1.04$ ($0.05$) \parbox[c]{5mm}{\centering$\rightarrow$}&$1.06$ ($0.34$) \parbox[c]{5mm}{\centering$\circ$}&\cellcolor{Gray}$1.03$ ($0.05$) \parbox[c]{5mm}{\centering$\rightarrow$}\\
Colombia (2013) & $1.25$ ($0.07$) \parbox[c]{5mm}{\centering$\nearrow$}&\cellcolor{Gray}\cellcolor{Gray2}$1.23$ ($0.07$) \parbox[c]{5mm}{\centering$\nearrow^*$}&$1.89$ ($0.09$) \parbox[c]{5mm}{\centering$\nearrow$}&\cellcolor{Gray}$1.31$ ($0.08$) \parbox[c]{5mm}{\centering$\nearrow$}\\
Colombia (2014) & $1.26$ ($0.07$) \parbox[c]{5mm}{\centering$\nearrow$}&\cellcolor{Gray}\cellcolor{Gray2}$1.24$ ($0.09$) \parbox[c]{5mm}{\centering$\nearrow^*$}&$1.89$ ($0.09$) \parbox[c]{5mm}{\centering$\nearrow$}&\cellcolor{Gray}$1.36$ ($0.08$) \parbox[c]{5mm}{\centering$\nearrow$}\\
Denmark (2015) & \cellcolor{Gray}\cellcolor{Gray2}$1.28$ ($0.10$) \parbox[c]{5mm}{\centering$\nearrow^*$}&$1.27$ ($0.13$) \parbox[c]{5mm}{\centering$\nearrow^*$}&$1.45$ ($0.33$) \parbox[c]{5mm}{\centering$\nearrow$}&\cellcolor{Gray}$1.27$ ($0.14$) \parbox[c]{5mm}{\centering$\nearrow$}\\
Denmark (2016) & \cellcolor{Gray}\cellcolor{Gray2}$1.27$ ($0.14$) \parbox[c]{5mm}{\centering$\nearrow^*$}&$1.28$ ($0.18$) \parbox[c]{5mm}{\centering$\nearrow^*$}&$1.58$ ($0.37$) \parbox[c]{5mm}{\centering$\nearrow$}&\cellcolor{Gray}$1.28$ ($0.18$) \parbox[c]{5mm}{\centering$\nearrow$}\\
France (2013) & \cellcolor{Gray}\cellcolor{Gray2}$1.24$ ($0.09$) \parbox[c]{5mm}{\centering$\nearrow$}&$1.23$ ($0.07$) \parbox[c]{5mm}{\centering$\nearrow^*$}&$1.59$ ($0.44$) \parbox[c]{5mm}{\centering$\nearrow$}&\cellcolor{Gray}$1.30$ ($0.12$) \parbox[c]{5mm}{\centering$\nearrow$}\\
France (2014) & $1.24$ ($0.10$) \parbox[c]{5mm}{\centering$\nearrow$}&\cellcolor{Gray}\cellcolor{Gray2}$1.22$ ($0.08$) \parbox[c]{5mm}{\centering$\nearrow$}&$1.70$ ($0.57$) \parbox[c]{5mm}{\centering$\nearrow$}&\cellcolor{Gray}$1.34$ ($0.18$) \parbox[c]{5mm}{\centering$\nearrow$}\\
Italy (2014) & \cellcolor{Gray}\cellcolor{Gray2}$1.33$ ($0.11$) \parbox[c]{5mm}{\centering$\nearrow^*$}&$1.31$ ($0.10$) \parbox[c]{5mm}{\centering$\nearrow^*$}&$1.37$ ($0.15$) \parbox[c]{5mm}{\centering$\nearrow$}&\cellcolor{Gray}$1.31$ ($0.09$) \parbox[c]{5mm}{\centering$\nearrow$}\\
Italy (2015) & \cellcolor{Gray}\cellcolor{Gray2}$1.32$ ($0.09$) \parbox[c]{5mm}{\centering$\nearrow^*$}&$1.29$ ($0.11$) \parbox[c]{5mm}{\centering$\nearrow^*$}&$1.35$ ($0.14$) \parbox[c]{5mm}{\centering$\nearrow$}&\cellcolor{Gray}$1.29$ ($0.10$) \parbox[c]{5mm}{\centering$\nearrow$}\\
Mexico (2015) & \cellcolor{Gray}\cellcolor{Gray2}$1.30$ ($0.04$) \parbox[c]{5mm}{\centering$\nearrow$}&$1.31$ ($0.04$) \parbox[c]{5mm}{\centering$\nearrow$}&$1.98$ ($0.02$) \parbox[c]{5mm}{\centering$\nearrow$}&\cellcolor{Gray}$1.32$ ($0.04$) \parbox[c]{5mm}{\centering$\nearrow$}\\
Mexico (2016) & \cellcolor{Gray}\cellcolor{Gray2}$1.26$ ($0.04$) \parbox[c]{5mm}{\centering$\nearrow$}&$1.26$ ($0.04$) \parbox[c]{5mm}{\centering$\nearrow$}&$1.98$ ($0.01$) \parbox[c]{5mm}{\centering$\nearrow$}&\cellcolor{Gray}$1.30$ ($0.05$) \parbox[c]{5mm}{\centering$\nearrow$}\\
South Africa (2016) & \cellcolor{Gray}\cellcolor{Gray2}$0.97$ ($0.11$) \parbox[c]{5mm}{\centering$\rightarrow^*$}&$0.99$ ($0.10$) \parbox[c]{5mm}{\centering$\rightarrow^*$}&$1.33$ ($0.20$) \parbox[c]{5mm}{\centering$\nearrow$}&\cellcolor{Gray}$1.02$ ($0.11$) \parbox[c]{5mm}{\centering$\rightarrow$}\\
Spain (2015) & $1.18$ ($0.11$) \parbox[c]{5mm}{\centering$\nearrow$}&\cellcolor{Gray}\cellcolor{Gray2}$1.19$ ($0.11$) \parbox[c]{5mm}{\centering$\nearrow$}&$1.27$ ($0.19$) \parbox[c]{5mm}{\centering$\nearrow$}&\cellcolor{Gray}$1.22$ ($0.12$) \parbox[c]{5mm}{\centering$\nearrow$}\\
Spain (2016) & $1.20$ ($0.11$) \parbox[c]{5mm}{\centering$\nearrow$}&\cellcolor{Gray}\cellcolor{Gray2}$1.20$ ($0.11$) \parbox[c]{5mm}{\centering$\nearrow$}&$1.31$ ($0.20$) \parbox[c]{5mm}{\centering$\nearrow$}&\cellcolor{Gray}$1.24$ ($0.13$) \parbox[c]{5mm}{\centering$\nearrow$}\\
United Kingdom (2015) & $1.24$ ($0.07$) \parbox[c]{5mm}{\centering$\nearrow$}&\cellcolor{Gray}\cellcolor{Gray2}$1.31$ ($0.10$) \parbox[c]{5mm}{\centering$\nearrow$}&$1.45$ ($0.30$) \parbox[c]{5mm}{\centering$\nearrow$}&\cellcolor{Gray}$1.55$ ($0.32$) \parbox[c]{5mm}{\centering$\nearrow$}\\
United Kingdom (2016) & $1.26$ ($0.09$) \parbox[c]{5mm}{\centering$\nearrow$}&\cellcolor{Gray}\cellcolor{Gray2}$1.33$ ($0.10$) \parbox[c]{5mm}{\centering$\nearrow$}&$1.50$ ($0.37$) \parbox[c]{5mm}{\centering$\nearrow$}&\cellcolor{Gray}$1.59$ ($0.35$) \parbox[c]{5mm}{\centering$\nearrow$}\\
United States (2014) & $1.12$ ($0.01$) \parbox[c]{5mm}{\centering$\nearrow$}&\cellcolor{Gray}\cellcolor{Gray2}$1.06$ ($0.01$) \parbox[c]{5mm}{\centering$\nearrow$}&$1.07$ ($0.06$) \parbox[c]{5mm}{\centering$\nearrow$}&\cellcolor{Gray}$1.04$ ($0.04$) \parbox[c]{5mm}{\centering$\nearrow$}\\
United States (2015) & $1.13$ ($0.01$) \parbox[c]{5mm}{\centering$\nearrow$}&\cellcolor{Gray}\cellcolor{Gray2}$1.06$ ($0.01$) \parbox[c]{5mm}{\centering$\nearrow$}&$1.08$ ($0.07$) \parbox[c]{5mm}{\centering$\nearrow$}&\cellcolor{Gray}$1.05$ ($0.04$) \parbox[c]{5mm}{\centering$\nearrow$}\\
\botrule
    \end{tabular}
\end{table}
\begin{table}[t!]
\centering
	\caption{$\beta$ estimates for the case of burglaries using log-normal and normal fluctuations.\label{tab:results_burglary}}
	\def\arraystretch{1.1}
    \setlength{\tabcolsep}{1.5pt}
    \begin{tabular}{r|ll|ll}
    \toprule
    \multicolumn{1}{c|}{}&\multicolumn{2}{c|}{Log-normal} & \multicolumn{2}{c}{Gaussian} \\
    & \multicolumn{1}{c}{$\delta =2$} & \multicolumn{1}{c|}{$\delta \in [1, 3]$} & \multicolumn{1}{c}{$\delta = 1$} & \multicolumn{1}{c}{$\delta \in [1, 2]$} \\\hline
  Belgium (2015) & $1.10$ ($0.06$) \parbox[c]{5mm}{\centering$\nearrow$}&\cellcolor{Gray}\cellcolor{Gray2}$1.09$ ($0.05$) \parbox[c]{5mm}{\centering$\circ$}&$1.21$ ($0.11$) \parbox[c]{5mm}{\centering$\nearrow$}&\cellcolor{Gray}$1.09$ ($0.05$) \parbox[c]{5mm}{\centering$\circ$}\\
Belgium (2016) & $1.08$ ($0.06$) \parbox[c]{5mm}{\centering$\circ$}&\cellcolor{Gray}\cellcolor{Gray2}$1.07$ ($0.05$) \parbox[c]{5mm}{\centering$\circ$}&$1.18$ ($0.10$) \parbox[c]{5mm}{\centering$\nearrow$}&\cellcolor{Gray}$1.08$ ($0.05$) \parbox[c]{5mm}{\centering$\circ$}\\
Canada (2015) & $0.93$ ($0.05$) \parbox[c]{5mm}{\centering$\circ$}&\cellcolor{Gray}\cellcolor{Gray2}$0.93$ ($0.04$) \parbox[c]{5mm}{\centering$\searrow^*$}&$1.04$ ($0.10$) \parbox[c]{5mm}{\centering$\rightarrow$}&\cellcolor{Gray}$0.95$ ($0.06$) \parbox[c]{5mm}{\centering$\circ$}\\
Canada (2016) & $0.91$ ($0.04$) \parbox[c]{5mm}{\centering$\searrow^*$}&\cellcolor{Gray}\cellcolor{Gray2}$0.90$ ($0.05$) \parbox[c]{5mm}{\centering$\searrow^*$}&$1.00$ ($0.10$) \parbox[c]{5mm}{\centering$\rightarrow$}&\cellcolor{Gray}$0.90$ ($0.04$) \parbox[c]{5mm}{\centering$\searrow$}\\
Colombia (2013) & $0.90$ ($0.07$) \parbox[c]{5mm}{\centering$\circ^*$}&\cellcolor{Gray}\cellcolor{Gray2}$0.93$ ($0.07$) \parbox[c]{5mm}{\centering$\rightarrow$}&$1.18$ ($0.44$) \parbox[c]{5mm}{\centering$\rightarrow$}&\cellcolor{Gray}$0.96$ ($0.07$) \parbox[c]{5mm}{\centering$\rightarrow$}\\
Colombia (2014) & $0.94$ ($0.07$) \parbox[c]{5mm}{\centering$\rightarrow^*$}&\cellcolor{Gray}\cellcolor{Gray2}$0.95$ ($0.06$) \parbox[c]{5mm}{\centering$\rightarrow$}&$1.16$ ($0.51$) \parbox[c]{5mm}{\centering$\rightarrow$}&\cellcolor{Gray}$0.99$ ($0.07$) \parbox[c]{5mm}{\centering$\rightarrow$}\\
Denmark (2015) & $1.11$ ($0.26$) \parbox[c]{5mm}{\centering$\rightarrow$}&\cellcolor{Gray}$0.91$ ($0.14$) \parbox[c]{5mm}{\centering$\rightarrow$}&\cellcolor{Gray}\cellcolor{Gray2}$0.92$ ($0.14$) \parbox[c]{5mm}{\centering$\rightarrow^*$}&$0.93$ ($0.13$) \parbox[c]{5mm}{\centering$\rightarrow^*$}\\
Denmark (2016) & $1.15$ ($0.24$) \parbox[c]{5mm}{\centering$\rightarrow$}&\cellcolor{Gray}\cellcolor{Gray2}$0.89$ ($0.15$) \parbox[c]{5mm}{\centering$\rightarrow$}&\cellcolor{Gray}$0.90$ ($0.13$) \parbox[c]{5mm}{\centering$\rightarrow$}&$0.92$ ($0.17$) \parbox[c]{5mm}{\centering$\rightarrow$}\\
France (2013) & \cellcolor{Gray}\cellcolor{Gray2}$1.29$ ($0.09$) \parbox[c]{5mm}{\centering$\nearrow^*$}&$1.27$ ($0.09$) \parbox[c]{5mm}{\centering$\nearrow^*$}&$1.31$ ($0.11$) \parbox[c]{5mm}{\centering$\nearrow^*$}&\cellcolor{Gray}$1.27$ ($0.09$) \parbox[c]{5mm}{\centering$\nearrow$}\\
France (2014) & \cellcolor{Gray}\cellcolor{Gray2}$1.29$ ($0.10$) \parbox[c]{5mm}{\centering$\nearrow^*$}&$1.27$ ($0.10$) \parbox[c]{5mm}{\centering$\nearrow^*$}&$1.34$ ($0.10$) \parbox[c]{5mm}{\centering$\nearrow$}&\cellcolor{Gray}$1.27$ ($0.09$) \parbox[c]{5mm}{\centering$\nearrow$}\\
Italy (2014) & \cellcolor{Gray}\cellcolor{Gray2}$1.13$ ($0.15$) \parbox[c]{5mm}{\centering$\rightarrow^*$}&$1.11$ ($0.16$) \parbox[c]{5mm}{\centering$\rightarrow^*$}&$1.09$ ($0.17$) \parbox[c]{5mm}{\centering$\rightarrow$}&\cellcolor{Gray}$1.09$ ($0.12$) \parbox[c]{5mm}{\centering$\rightarrow^*$}\\
Italy (2015) & \cellcolor{Gray}\cellcolor{Gray2}$1.09$ ($0.15$) \parbox[c]{5mm}{\centering$\rightarrow^*$}&$1.07$ ($0.13$) \parbox[c]{5mm}{\centering$\rightarrow^*$}&$1.06$ ($0.15$) \parbox[c]{5mm}{\centering$\rightarrow^*$}&\cellcolor{Gray}$1.05$ ($0.12$) \parbox[c]{5mm}{\centering$\rightarrow^*$}\\
Portugal (2015) & \cellcolor{Gray}\cellcolor{Gray2}$0.99$ ($0.06$) \parbox[c]{5mm}{\centering$\rightarrow^*$}&$0.98$ ($0.05$) \parbox[c]{5mm}{\centering$\rightarrow^*$}&$1.13$ ($0.13$) \parbox[c]{5mm}{\centering$\rightarrow$}&\cellcolor{Gray}$0.99$ ($0.10$) \parbox[c]{5mm}{\centering$\rightarrow$}\\
Portugal (2016) & \cellcolor{Gray}\cellcolor{Gray2}$1.02$ ($0.05$) \parbox[c]{5mm}{\centering$\rightarrow$}&$1.01$ ($0.06$) \parbox[c]{5mm}{\centering$\rightarrow$}&$1.11$ ($0.09$) \parbox[c]{5mm}{\centering$\circ$}&\cellcolor{Gray}$1.05$ ($0.10$) \parbox[c]{5mm}{\centering$\rightarrow$}\\
South Africa (2016) & \cellcolor{Gray}\cellcolor{Gray2}$0.91$ ($0.09$) \parbox[c]{5mm}{\centering$\rightarrow$}&$0.91$ ($0.08$) \parbox[c]{5mm}{\centering$\rightarrow$}&$1.07$ ($0.09$) \parbox[c]{5mm}{\centering$\circ$}&\cellcolor{Gray}$0.97$ ($0.12$) \parbox[c]{5mm}{\centering$\rightarrow$}\\
United Kingdom (2015) & \cellcolor{Gray}\cellcolor{Gray2}$1.39$ ($0.11$) \parbox[c]{5mm}{\centering$\nearrow^*$}&$1.42$ ($0.10$) \parbox[c]{5mm}{\centering$\nearrow^*$}&$1.47$ ($0.13$) \parbox[c]{5mm}{\centering$\nearrow$}&\cellcolor{Gray}$1.40$ ($0.10$) \parbox[c]{5mm}{\centering$\nearrow$}\\
United Kingdom (2016) & \cellcolor{Gray}\cellcolor{Gray2}$1.35$ ($0.11$) \parbox[c]{5mm}{\centering$\nearrow^*$}&$1.36$ ($0.10$) \parbox[c]{5mm}{\centering$\nearrow^*$}&$1.46$ ($0.14$) \parbox[c]{5mm}{\centering$\nearrow$}&\cellcolor{Gray}$1.37$ ($0.11$) \parbox[c]{5mm}{\centering$\nearrow$}\\
United States (2014) & $0.99$ ($0.01$) \parbox[c]{5mm}{\centering$\rightarrow$}&\cellcolor{Gray}\cellcolor{Gray2}$0.99$ ($0.01$) \parbox[c]{5mm}{\centering$\rightarrow$}&$1.19$ ($0.11$) \parbox[c]{5mm}{\centering$\nearrow$}&\cellcolor{Gray}$1.07$ ($0.05$) \parbox[c]{5mm}{\centering$\nearrow$}\\
United States (2015) & \cellcolor{Gray}\cellcolor{Gray2}$0.98$ ($0.01$) \parbox[c]{5mm}{\centering$\rightarrow$}&$0.98$ ($0.01$) \parbox[c]{5mm}{\centering$\circ$}&$1.17$ ($0.08$) \parbox[c]{5mm}{\centering$\nearrow$}&\cellcolor{Gray}$1.07$ ($0.06$) \parbox[c]{5mm}{\centering$\nearrow$}\\
\botrule
    \end{tabular}
\end{table}

\subsection*{Appendix II: Data sources}

\begin{table}[h!]
\centering
	\caption{Data sources for each country.}
   \renewcommand{\arraystretch}{1.3}
   \setlength{\tabcolsep}{4pt}
\begin{tabular}{lp{4.5cm}}
\toprule
Country & URL \\\hline
Belgium        & \url{https://bit.ly/3v5I8am} \\
Canada         & \url{https://bit.ly/3f42Jq7} \\
Colombia       & \url{https://bit.ly/3u7lFZb} \\
Denmark        & \url{https://bit.ly/3bETREY} \\
France         & \url{https://bit.ly/3u8727Q} \\
Italy          & \url{https://bit.ly/2S9iAL0} \\
Mexico         & \url{https://bit.ly/3v5JFgC}  \url{https://bit.ly/3lkch3w} \\
Portugal       & \url{https://bit.ly/3hHbMz0} \\
South Africa   & \url{https://bit.ly/3u7m6CN} \\
Spain          & \url{https://bit.ly/3wqOWjb} \\
United Kingdom & \url{https://bit.ly/3ytMOJ0} \\
United States  & \url{https://bit.ly/3fx2iTZ} \\
\botrule    
    \end{tabular}
\end{table}

\end{document}